\documentclass[12pt]{article}

 \usepackage{amsfonts}   \usepackage{amssymb,amsmath}
 \usepackage{graphicx} \usepackage{epstopdf}
 \newcommand{\crlb}[1]{\label{#1}\\[2pt]}
 
 \newcommand{\crld}[1]{\label{#1}}
 \newcommand{\eela}[1]{\quad\hbox{\scriptsize{#1}}\label{#1}\end{eqnarray}}
 \newcommand{\eelb}[1]{\label{#1}\end{eqnarray}}
 
 \newcommand{\newsecb}[2]{\section{#1}\label{#2}\setcounter{equation}{0}}

 \newcommand{\nolabels} {\def\eel{\eelb} \def\crl{\crlb} \def\newsecl{\newsecb}\def\bibiteml{\bibitem}\def\citel{\cite}\def\labell{\crld}}

\newcommand\publishversion{\nolabels\setlength{\textheight}{8.75in}\setlength{\oddsidemargin}{0in}
    \setlength{\textwidth}{6.3in}\setlength{\topmargin}{-0.3in}}

                 \def\fn{\footnote}
     \def\nm{\nonumber}   \def\be{\begin{eqnarray}}    \def\ee{\end{eqnarray}}
 \def\bi#1{\begin{itemize}\item[#1]}   \def\itm#1{\item[#1]}  \def\ei{\end{itemize}}  \def\eqn#1{(\ref{#1})}
   \def\^#1{\hat{#1}}

          \def\g{\gamma}      \def\G{\Gamma}
 \def\d{\delta}        \def\e{\varepsilon} 
        \def\L{\Lambda}     \def\m{\mu}
 \def\f{\phi}        \def\F{\Phi}    \def\vv{\varphi}    \def\n{\nu}
 \def\j{\psi}                 \def\s{\sigma}  
         \def\tht{\theta}  
               
 \def\w{\omega}        

    \def\LL{{\mathcal L}} \def\OO{{\mathcal O}}  
 \def\pa{\partial} \def\ra{\rightarrow}
  
 \def\dd{{\rm d}}

 \def\fract#1#2{{\textstyle{#1\over#2}}}
 \def\ffract#1#2{\raise .2 em\hbox{$\scriptstyle#1$}\kern-.3em/
                 \kern-.2em\lower .15 em \hbox{$\scriptstyle#2$}}
 
 \def\half{\fract12} \def\quart{\fract14}

\def\bmatrix{\begin{matrix}} \def\ematrix{\end{matrix}} \def\bpmatrix{\begin{pmatrix}}\def\epmatrix{\end{pmatrix}}
\def\bcenter{\begin{center}} \def\ecenter{\end{center}}

\def\lowerheightfig#1#2#3{\(\raise-#1\hbox{\includegraphics[height=#2]{#3}}\)}
\def\lowerwidthfig#1#2#3{\(\raise-#1\hbox{\includegraphics[width=#2]{#3}}\)}

 \publishversion

\begin{document}

\bcenter 
{\LARGE\textbf{Reflections on the renormalization procedure for gauge theories\\[25pt]}}
{\large Gerard 't~Hooft}  \\[20pt]
Institute for Theoretical Physics \\[5pt]
\(\mathrm{EMME}\F\)   \\
Centre for Extreme Matter and Emergent Phenomena\\[5pt] 
Science Faculty\\ 
Utrecht University\\[5pt]
 POBox 80.195 \\
 3808TD, Utrecht  \\
The Netherlands  \\[10pt] 
{\small Internet:  http://www.staff.science.uu.nl/\~{}hooft101/ }
\vfil
In memory of Raymond Stora\\
1930 -- 2015
\ecenter
\vfil
\begin{quotation} \noindent {\large\bf Abstract } \medskip \\
	Various pieces of insight were needed to formulate the rules for working with gauge theories of the electro-magnetic, weak and strong forces.
	First, it was needed to understand how to formulate the Feynman rules. We had to learn that there are many different ways to derive them, and
	it was needed to know how different formulations of the gauge constraint lead to the same final results: the calculated values of the scattering amplitudes.
	The rules for dealing with the infinities that had to be subtracted were a big challenge, culminating in the discovery of the Becchi-Rouet-Stora-Tyutin
	symmetry. Fond recollections of the numerous discussions the author had with Raymond Stora on this topic are memorised here.	
	We end with some reflections on the mathematical status of quantum field theories, and a letter sent by Stora to the author.

\end{quotation}
\vfil 
\noindent Version: March 2016 \   \\[2pt]
{\footnotesize Last typeset: \today}
\eject
\setcounter{page}{2} 

\newsecl{Introduction}{intro}

	Around the year 1970, rapid changes took place in our views of elementary particle physics, and in particular the role of local gauge theories and spontaneous symmetry breaking in these theories. For a long time, only few researchers had been convinced that quantum field theory was the way to go. Handling strongly interacting hadronic particles was still quite tricky, but we were coming into grips with the perturbation expansion of these theories, when the interactions are sufficiently weak. 
	
	This seemed to be a well-posed problem, but for a long time the situation had been quite confuse. One researcher who had been quite determined to crack the problem was my teacher and advisor M.~Veltman. Veltman had concluded from experimental evidence that the gauge theory of Yang and Mills had to contain a lot of truth. Eventually, we succeeded together. The key to the answer was that local gauge invariance may never be explicitly broken, particularly not by the interactions, so that all gauge boson masses, and the masses of many of the fermions, had to be attributed to what is now called the Brout-Englert-Higgs mechanism.
	 
	To set up a model, and to do calculations with it, several steps had to be taken:
	\bi{-} One had to guess the symmetry algebra. There were local and global gauge symmetries. Global symmetries may either be exact (such as -- in practice\fn{This symmetry turns out to be broken as well, albeit extremely weakly, by instanton effects.} -- baryon number conservation), or approximate (such as isospin and flavor SU(3)). It was now understood that the local symmetries may never be broken explicitly by any of the interactions, but one may invoke a spontaneous symmetry breaking mechanism. Strictly speaking, this does not break a local symmetry at all, it was just a procedure to do perturbation expansions around some non symmetric classical background configuration.
	\itm{-} Choose the various interaction constants, including the ones that give masses to particles such as the weak interaction vector bosons, and the various types of leptons. All those parameters would be found to be affected by renormalization, after the interactions have been computed.
	\itm{-} Choose how to fix the gauge freedom. This gauge-fixing should have no effect on the physical phenomena predicted by the model, but the way we do the calculations would depend heavily on this choice. Each choice of gauge fixing leads to different sets of Feynman rules. It's sort of a sport to find the most efficient gauge fixing choice and the associated Feynman rules to calculate something. The Feynman rules refer to Feynman diagrams. Each contribution to the calculation of something in perturbation theory can be characterised by a diagram.
	\itm{-} Calculate the diagrams, and discover that they lead to divergent integrals. By noticing that one had some freedom anyway in choosing the masses and coupling parameters of the theory, one discovers that all divergent expressions can be absorbed in redefinitions of these parameters. If all is done well, one obtains finite, meaningful expressions. Now to do this correctly, one first had to choose how to modify the theory ever so slightly such that all integrals do come out finite, even if the underlying theory now may violate concepts such as unitarity. One uses a small parameter \(\e\) to denote the modification. After the divergent parts have been arranged to cancel, one can calculate the limit \(\e\ra 0\). If this is done well, finite, meaningful expressions come out.
	\itm{-} Check that the resulting theory does not violate unitarity. This was not so easy, because during the intermediate steps, where \(\e\) was not yet taken equal to zero, unitarity was violated. It turned out that one must ensure that local gauge invariance is kept intact.
	\itm{-} Find a way to check what local gauge invariance means for these Feynman diagrams, and check whether the procedures used are mathematically sound. \ei

The author had numerous discussions about these points with Raymond Stora. Raymond was not happy with the fact that we were using and manipulating Feynman diagrams to see what the rules are, and to prove that these rules do what we expect them to do. Veltman and I thought that these Feynman diagrams can be handled just as the terms in a long equation, except that they show much better what the structures of these equations were. Stora was unhappy with that. In addition, Veltman and I had developed procedures called `dimensional regularization' and `dimensional renormalization'. We thought these approaches did exactly what we needed, that is, they provided unambiguous calculation prescriptions and proofs that these lead to the desired results, but Raymond insisted that we should look for more physical arguments.

\newsecl{Slavnov Taylor identities}{SlavnovTaylor}
An essential ingredient in our procedures proving the renormalizability of gauge theories, was that there are intimate relationships between the numerous Feynman diagrams that we encountered. The diagrams represented the contributions of all sorts of virtual particles, ensuring that the scattering amplitudes we obtained should be unitary. These scattering amplitudes obey dispersion relations: the imaginary parts of a scattering matrix \(S\) are related to the contributions of all virtual particles to the product \(S\cdot S^\dag\). In ordinary theories, \emph{i.e.} theories without non-Abelian gauge fields, one could recognise the desired features immediately as properties of the Feynman diagrams, but in the non-Abelian case, problems seemed to emerge: after fixing the gauge choice, the diagrams seemed to reflect the presence of unphysical particles, called `ghosts'. There seemed to be contributions of amplitudes producing ghost particles whereas one had to assume that those ghosts will never be observable in experiments. We had to face `negative probabilities'. In fact, we had to prove that all diagrams producing ghost particles should cancel out in the dispersion relations.

In my thesis, I did this by hand; we checked all diagrams and noticed which properties were needed for the cancellation. The cancellations had to occur when all particles, including the ghost particles, obeyed the mass shell condition. The amplitudes for all initial and final particles, in particular also the ghosts, had to cancel out. This we could prove, by finding crucial generalisations of the so-called Ward Takahashi identities\,\cite{Ward}\cite{Takahashi}, provided contributions called `anomalies' would cancel out. The anomalies would signal that local gauge invariance is broken somewhere. We thought we had this situation under control, but Stora was not content with our results. It should be done more elegantly. 

We had found that our identities for on-shell diagrams suffice to prove the internal consistence of the theory. To \emph{prove} these identities, we also needed properties of diagrams off-shell, but to avoid discussions of renormalizing the new infinities that would appear in such diagrams, we avoided to mention these explicitly. That was a mistake. Soon after our work on these points was published, two preprints appeared in the mail, one by A.A.~Slavnov\,\cite{Slavnov} and one by J.C.~Taylor\,\cite{Taylor}, both containing the same message: our identities among diagrams could be formulated off-shell as well. They were the first two authors who referred to my work. It so happened that Stora needed exactly their equations for understanding the internal properties of the theory. He coined the name `Slavnov-Taylor identities'. If there are infinities in the theory, one has to make subtractions to replace them by finite expressions. These subtractions should not violate gauge invariance.

Now the question remained whether possible anomalies could pop up. If you make your subtractions in one diagram, gauge invariance -- read Slavnov-Taylor identities -- should fix the expression for other diagrams. Will there be any clashes? Examples of such clashes were known: the \emph{Adler-Bell-Jackiw anomalies}\,\cite{ABJ}. These anomalies refer primarily to diagrams with a single, triangular fermion loop in them. For each diagram, they imply three constraints, each on one of the end points of the triangle, but there were only two parameters to fix. This gives us more equations than unknowns, and sometimes there is no solution. The physical reason for this clash would become evident much later (instantons), but just for now, we needed to know whether, and if so, when, there would be problems of this sort, interfering with the renormalization program. 

In passing, the anomaly is usually named after Adler, Bell and Jackiw, whose papers contain modern descriptions of the effect and were well-known in the western world, the \emph{discovery} of this anomaly goes back to Fukuda and Miyamoto, in a little-known paper written in 1949\,\cite{FM}.

For some time, Stora was experimenting with what was called `normal product renormalizations'. At some values of the external momenta, you impose a diagram to vanish. At which values? The most natural answer would be: have the diagrams vanish when all external momenta are zero. Then, the procedure would be unambiguous.

Whether that would work, unfortunately, was not obvious at all: \emph{Theories with local gauge-invariance, before the symmetry is spontaneously broken, have infra-red divergences}. This means that new infinities arise when the external momenta vanish, and these infinities should \emph{not} be subtracted. And if you have the Brout-Englert-Higgs (BEH) mechanism? Then there are no infra-red divergences. There could be massless ghosts, but by carefully picking the gauge -- in fact, Stora was particularly happy with the `\,'t Hooft gauge' -- there are no massless ghosts either. However, when the symmetry is spontaneously broken, the vacuum itself is not gauge-invariant. He soon found out that this means that normal product subtractions also fail in the BEH case.

Veltman and I had what we considered to be the answer to this question: \emph{dimensional regularization and renormalization}. This answer shows that there are cases where anomalies may impair the procedure: \emph{chiral} fermions cannot be generalised to higher (or lower) dimensions because of ambiguities in the definition of the Dirac matrix \(\g^5\). Indeed, it was found that, more often than not, such theories cannot be repaired. These are exactly the chiral theories where Adler, Bell and Jackiw had discovered their anomalies. One must require an explicit restriction on the algebra of chiral fermions.

\newsecl{BRST symmetry} {BRST}
	The idea that the Slavnov-Taylor identities for diagrams must be associated to a global symmetry of the system, had been considered by us, but we had dismissed it: symmetry generators obey Jacobi identities; the identities that were crucial for our diagrams did look like Jacobi identities, but unfortunately, the signs were all wrong. We had treated the diagrams just as if we were handling a symmetry, but always had to insert the signs in a way that was different. This, however, was exactly what Stora was looking for. By a stroke of genius, he, together with his co-authors A.~Rouet and C.~Becchi\,\cite{BRS}, and independently, as it turned out later, I.V.~Tyutin\,\cite{Tyutin}, discovered that there indeed \emph{was} a symmetry, now called BRST symmetry, that does the job impeccably, but it is a \emph{super}symmetry, a symmetry generated by an anti-commutator algebra instead of ordinary symmetries that use commutators.\fn{For the record, the old papers by B.S.~DeWitt on gravity\,\cite{DeWitt} discuss transformations very similar to BRST. Very shortly after the BRS publication, there were lecture notes of J.\ Zinn-Justin\,\cite{ZinnJustin} on the same topic.}.	
	 This explained our earlier problems with signs. Supersymmetry had been discovered earlier, but the idea that such a symmetry, normally only imposed on models with fermions, could apply to pure gauge theories as well, was novel, and great. \def\inv{{\mathrm{inv}}}\def\ol#1{\overline{#1}} \def\gf{{\mathrm{gaugefix}}}  \def\fp{{\mathrm{FP}}} \def\scrA{{\mathcal {A}}}

BRST is a global symmetry of the gauge theory Lagrangian, \emph{after} gauge fixing terms and ghost terms are added to this Lagrangian. When counter terms are added for renormalization purposes, all that was needed was to check that these are invariant under the BRST transformation. From that point on, the Feynman diagrams are no longer needed as essential ingredients for the formulation of the theory, but they can be put where they belong: in our bags of machinery to compute amplitudes. this implies a considerable streamlining of our procedures. 

It is a characteristic feature in many branches of science, in particular mathematical physics, that the first results that are discovered in a process to understand the physical world, are complicated, and difficult to teach, until someone comes along who states that it all can be done much simpler. This is what Becchi, Rouet, Stora and Tyutin succeeded in doing.

BRST symmetry is usually introduced as follows. In a short-hand notation, the Lagrangian on which a gauge theory is based, can be written as
	\be\LL_\inv&=&-\quart F_{\m\n}F_{\m\n}-\ol{\j}(\g D+m+W(\f))\j-\half (D_\m\f)^2-V(\f)\ ,\nm\\
	 F_{\m\n}&\equiv& \pa_\m A_\n-\pa_\n A_\m+g [A_\m,A_\n]\ , \quad W(\f)=W_1(\f)+i\g^5W_2(\f)\ .\eel{Linv}
This must be invariant under a local gauge transformation, which, in its infinitesimal form appears as (also in short-hand):
	\be A_\m(x)&\ra&A_\m(x) -D_\m \L(x)\ ,\nm \\
		\f(x)&\ra&\f(x)+i\L(x)\f(x)\ ,\nm \\		\j(x)&\ra&\j(x)+i\L(x)\j(x)\ .	\ee
Fixing the gauge condition amounts to adding a gauge-fixing term to the Lagrangian:
	\be \LL_\gf = -\half B(A,\f)^2\ , \eel{gaugefix}
but, as was discovered by Feynman\,\cite{Feynman}, DeWitt\,\cite{DeWitt}, Faddeev and Popov\,\cite{FP}, the Feynman rules derived directly from this Lagrangian added to \eqn{Linv}, do not lead to unambiguous results, unlike the case in Maxwell theory, which is an \emph{Abelian} gauge theory. What was missing was a `ghost' term. If, under an infinitesimal gauge transformation, the field \(B(A,\f)\) transforms as
	\be B(x)\ra B(x)+\frac{\pa B(x)}{\pa\L^a(x')}\,\L^a(x')\ , \ee
then the Faddeev-Popov `ghost  Lagrangian' reads:
	\be \LL_\fp(x)=\ol\eta^a(x)\,\frac{\pa B^a(x)}{\pa\L^b(x')}\,\eta^b(x')\ ,\eel{faddeevpopov}
where the fields \(\eta\) and \(\ol\eta\) act as anticommuting fields, just as the fermion fields. \def\qqquad{\qquad\qquad}

The BRST transformation mixes the gauge fields with the ghost fields, and is  based on an anticommuting generator field \(\ol\e\).
One then can check that \( \LL_\inv+\LL_\gf+\LL_\fp \) is invariant under the global supersymmetry transformation:
	\begin{subequations} \labell{susy}
	\begin{align}
	\scrA(x)&\ra\qquad \scrA+\ol\e\,\frac{\pa \scrA(x)}{ \pa \L^a(x')} \,\eta^a(x')\ ;`& \\[2pt]
		\eta^a(x)&\ra \ \eta^a(x)+\half\ol\e f_{abc}\eta^b(x)\eta^c(x)\ ;  &\labell{etatrf}  \\[5pt]
		\ol\eta^a(x)&\ra\quad \ \ol\eta^a(x)+\ol\e B^a(x)\ , &\labell{etacancel} 
	\end{align}
	\end{subequations}
where \(\scrA\) stands for all fields that transform non-trivially under the local gauge transformations, and \(f_{abc}\) are the structure constants of the group.

In some of the original presentations, these notions were written in a mathematical language that was sometimes difficult to comprehend, and since we had found that the `Slavnov-Taylor' identities do \emph{not} refer to an ordinary symmetry, so that they were difficult to prove, I asked Stora whether he had to rely on Jacobi's identity for the structure constants of the group, 
	\be f_{pqa}\,f_{qbc}+f_{pqb}\,f_{qca}+f_{pqc}\,f_{qab} = 0 \, \eel{Jacobi}
which had been a crucial element of our proofs using the diagrams. ``\emph{Yes!}", was his reply, ``we also need this identity in our proofs". That was when I was assured that he must have done the right thing. 

In modern language, the invariance of the Lagrangian  \(\LL_\inv+\LL_\gf+\LL_\fp\) under the supersymmetry transformation \eqn{susy} is easy to check, except perhaps the variation of the Faddeev-Popov Lagrangian \(\LL^\fp\) against the contribution of Eq.\ \eqn{etatrf}: 
	\be\ol\eta^a\,\frac{\pa B^a}{\pa\L^b}\half\ol\e f_{bcd}\eta^c\eta^d+\ol\eta^a\frac{\pa}{\pa\L^c}\frac{\pa B^a}{\pa\L^d}\ol\e\eta^c\eta^d\ =\ \cdots\ . \eel{FPcancel}
Substituting some practical examples for the gauge constraint function \(B^a\), one discovers that these terms always cancel out. Of course, this is the Jacobi identity \eqn{Jacobi}, or more precisely, the statement that the commutator of two gauge transformation generators acting on the gauge-fixing term is again a generator of a gauge transformation.

\newsecl{The physical interpretation of BRST symmetry}{phys}
	When we gauge-fix a local symmetry, the local symmetry is supposed to disappear. One may wonder why it is that, in the same process, a supersymmetry appears. The symmetry one would really expect is a symmetry transforming from one gauge-fix formalism to another. Why is this a supersymmetry?
	
	To understand this, one has to realise that integration over Grassmann variables is very much like differentiation. The reason is this. Let there be a function \(f(x)=f(0)+x\,f'(0)+\OO(x^2)\). Now take \(\tht\) to be a Grassmann variable, with \(\tht^2=0\). When we have an  integral of the form
		\begin{subequations}\begin{align}  \int\dd\tht f(\tht)&=\int\dd\tht (f(0)+\tht\,f'(0))&\\
			&=f'(0)\ ,\qquad\hbox{since}\  \int\dd\tht=0\,,\ \int\dd\tht\,\tht\equiv 1\ ,&
		\end{align}\end{subequations}
so we end up with the differential. This is also reflected in the Gaussian integrals. Here, we see that a gaussian integral over an \(n\)-dimensional complex Grassmann variable is the inverse of the Gaussian integral of an ordinary complex vector in the \(n\)-dimensional complex plane:
		\be \int\dd^n\vec\vv\,\dd^n\vec\vv^{\,*}\,e^{-\half\vec \vv^{\,*} A\, \vec\vv}=(2\pi)^{-n}{\det}^{-1}(A)\ ;\quad\int\dd^n\vec\tht\,\dd^n\vec{\ol\tht} e^{-\half\vec{\ol\tht} A\vec\tht}=2^{-n}\det (-A) \eel{gaussians}
Thus, if we would have the product of a Gaussian integral over and ordinary variable \(\vv\) and a Grasmann variable \(\tht\), with the same matrix \(A\), their effects completely cancel out (apart from a constant). 

This is exactly why the Faddeev-Popov Lagrangian \(\LL_\fp\), Eq.\ \eqn{faddeevpopov} has a fermionic variable in it: it must cancel the effects of the integral over the gauge-fixing part of the total Lagrangian, when integrating over the gauge sectors.

Thus, if we wish to \emph{guarantee} that the Faddeev-Popov ghost restores independence of the gauge fixing Lagrangian \eqn{gaugefix}, we must ensure that, effectively, we have two Gaussian integrals whose determinants cancel out,  or, we want Eq.~\eqn{gaussians} to be 100\% effective. This works  if the integrand comes in the combination \(\vec\vv^{\,*}\vec\vv+\vec{\ol\tht}\vec\tht\), which is why we demand a symmetry transforming \(\vec\vv\) (the integration variable denoting gauge-sectors) into \(\vec\tht\), and \(\vec\vv^{\,*}\) into \(\vec{\ol{\tht}}\), and back. It is a supersymmetry.
	
\newsecl{How rigorous is quantum (gauge) field theory mathematically?}{rigor} \def\tth{\(\!^\mathrm{th}\)}
	Physics is a science based on precision and rigour. Quantum field theories, in particular gauge field theories are mathematically quite complex, and of course we feel the need to handle these theories as precisely as possible. What has been achieved, during most of the 20\tth century, is an accurate formulation of the Taylor expansion of quantum field theories with respect to all interaction parameters -- with in addition some exponential effects such as the ones caused by instantons. This means that, if \(\vec g\equiv\{g_1,\dotsc,g_n\}\) is the set of (dimensionless) renormalized interaction parameters, then all scattering amplitudes and decay amplitudes can be expressed as series up to any fixed order \(N\):
		\be \G(\vec g)=\sum\limits_{i_1+\dotsb+i_n<N}\G_{i_1,\dotsc,i_n}(\vec g)\,g_1^{i_1}\dotsm g_n^{i_n}+\OO(\vec g)^{N}\ . \ee
These first \(N-1\) coefficients obey all unitarity and positivity constraints one would like to see as if this were en exact asymptotic expansion of physically meaningful amplitudes. The problem, however, is to find constraints on the values that the higher order remainders might have. How divergent is this series?

Just because one of these field theories, the `Standard Model', agrees with experiment so well, physicists in practice do not expect great catastrophes that will offset this magnificent achievement. Quantum chromodynamics (QCD) in particular, can be handled using lattice simulations, and although precision that can be reached this way is not very high (presently at the order of 1\%), one gets the general impression that the actual predictive power of this theory is not subject to any bounds.

Such optimism, however, might not be justified. From several corners, one may be confronted with arguments saying that quantum field theories, even the asymptotically free ones, cannot by themselves lead to infinitely precise predictions. Mathematically, one generally finds that the \(\OO(N)\) terms typically diverge as 
	\be\OO(N)\ra C^N\,N!\,(\vec g)^N\ . \ee
Using Stirling, one finds that, if this is so, the most accurate value that can be obtained for an amplitude has a margin typically of the form
	\be |\d \G|=\OO(e^{-1/gC})\ . \eel{margin}
In most theories, the coupling constants \(g\) are so small that there is no reason to expect difficulties with the predictive power of the theory in practice, but this does mean that there are fundamental margins of uncertainty in principle. Can one improve the theory?

Margins of uncertainty come from many other sources as well: we do not know the actual values of the couplings and the masses very precisely, but there is also a more fundamental difficulty: there is no reason to take our present theories seriously at extremely high energies, most notably the energies beyond the Planck scale, where gravitational forces generate difficulties of a more fundamental nature. 

Yet, we can ask whether quantum field theories by themselves can have a rigorous mathematical basis at all. Consider quantum electrodynamics. At 
energies comparable to those of the Landau ghost, the theory becomes ill-defined. This leads to uncertainties comparable to Eq.~\eqn{margin}. The only conceivable cure for the theory would be to embed it in some grand unified system, which would replace the Landau ghosts by some decently behaved particles, with new physical degrees of freedom.

Many theorists believe that the situation is fundamentally better for \emph{asymptotically free} theories. These have no Landau ghosts at high energy, The low-energy behaviour is compounded by infra-red divergences that may lead to features such as permanent confinement as in QCD, but there is no fundamental problem with that. 

Fact is, however, that there is no proof of the mathematical existence of such a model beyond its perturbation expansion. We do not know of any such theory in 4 space-time dimensions, except in some \(N\ra\infty\) limit. This leads me to considering an other scenario.

\newsecl{Foundations of quantum field theory}{found}
The ultimate cure that might restore the mathematical existence of relativistic quantum mechanics might be the inclusion of the gravitational force. We then get \emph{general} relativistic quantum mechanics. Today, no rigorous theory of that kind exists either, but one might hope that, eventually, a sound theory of nature will be found. It must obviously include the gravitational force, besides quantum mechanics. The author has investigated theories that possess an ultimate smallest distance limit: the cellular automata\,\cite{GtHCA}.

Our claim is that a mapping may exist that turns certain cellular automaton theories into genuine quantum field theories. The underlying theory definitely exists mathematically, since it is completely classical and completely local. What is unknown today is whether such models can be persuaded to exhibit Lorentz invariance, let alone general coordinate invariance, but we know that, at least in the real world, nature managed to generate these symmetries. We  also know that our models may well generate quantum behaviour. The difficulty is to derive that the quantum models we can reproduce can correspond to anything resembling the Standard Model. Our conjecture is that this might be the case.

The only thing then that we do not know is whether a completely air-tight quantum field theory will be generated. The reason for mentioning this work here is now, that we have a mathematical procedure in mind. We may be able to create models that allow for a systematic perturbation expansion. The expansion parameter may be a freely adjustable parameter such as \(1/N\), where \(N\) is a large number specifying the symmetry algebra that we may use. Such an approach is entirely doable. Perturbation expansions are under control. At \(N\ra\infty\), our theory describes free particles behaving as quantum objects, and then some `rare events' in the cellular automaton may make these particles interact. The perturbation expansion can perhaps be arranged in such a way that it is mathematically identified with the perturbation expansion of a quantum field theory.

What we would then have is a theory that is not quantum mechanical at its core, but deviations from standard quantum mechanics would be hidden exactly at the same spots where today's quantum field theories may become inconsistent because of non-convergence of the perturbation expansion. The situation that we then arrive at would be that we have not achieved the ultimate, mathematically tight quantum field theory, but instead,  a theory that is as good as quantum field theory is today, while it now does have an air-tight description in terms of a cellular automaton at the smallest possible distance scale, a description of the kind that we theoretical physicists should really be pursuing today.

Needless to emphasise that the considerations of this last section are pure speculation, and the reader may take them for what they are.

\newsecl{On the historical events of the early 1970s}{hist}

In his beautiful handwriting, Raymond Stora wrote me a letter dated october 13, 1994, stating some of his recollections and asking me some questions. I am assuming he would not object against reproducing the letter here:
\pagebreak
{\flushright{Annecy Le Vieux, le October 13, 1994}\\[5pt]
}\noindent Dear Gerard,\\
Georges Girardi extracted from the machine your Erice talk entitled ``Gauge theory and renormalization"\cite{GtHErice1994} which I enjoyed reading. I like history and stories, when time allows. The Bell Treiman transformation is among the best I know!\fn{Veltman had confided to the author that he named the transformation he found to be useful, after Bell and Treiman, ``because these gentlemen get so many features in particle physics named after them!". The idea was to add a free field to the system, which then is mixed with the existing fields by means of a gauge transformation. ``Veltman transformation" would have been more appropriate, but neither the name, nor the transformation itself, were much used afterwards.}

I myself tried to reconstruct some history around the BRST story and found some haze on the dates, even with the help of my friends. I have some haze on other points concerning the 1971 and 1972 Marseille meetings, which you may remember better. At the time, I wanted to understand better K.\ Symanzik's 1970 paper on the \(\s\) model. I had had long discussions with him. -- that was a new way to think about renormalization, pre-BPHZ in some sense compared to using ``good regularizations" \& renormalization. It allowed to resum breaking corrections, provided one could prove that some Ward identity could be fulfilled to all orders (which he did not bother to do). I would call this ``stupid renormalization", which sounds crazy. I decided to try and test that method on gauge theories which looked ``interesting enough" due to the seemingly large amount of poison it contained.

From you I learnt how to go from the Landau gauge to the Feynman gauge (some Russian papers were totally opaque to me), in order to avoid unphysical massless phenomena. Next there was the question of finding the Ward identity -- I had spotted a strange trick in Slavnov's preprint who only applied it to the ghost antighost 2 point function. I was unaware of Taylor's papers. I remember I gave you a xerox of Slavnov's paper, in Marseille, presumably in '72, because his identity looked very much like your graphical identities.\fn{J.\,C.~Taylor had sent a preprint of his work to the author in 1971; I heard of Slavnov's work shortly after that, presumably when Stora gave me a xerox of his work, but this I do not quite remember.}

In 1973 I spent a year in CERN and talked a lot to Tini. I had to lecture in Lausanne and decided to do Slavnov's exercise in full, with the help of Claude Itzykson. I send you the corresponding pages. I also met Carlo Becchi \& invited him to come to Marseille in 1973 -- 1974. After finishing some papers, he decided to read the Lausanne notes and fell on Eq.~8 p.~50. He jumped to Rouet's office (Rouet was working on his thesis) and said: Eq.~8 means the Lagrangian has an invariance, since 8 is linear in the sources -- Alain read off the answer. I came back the next day \& clapped my hands -- They say this was around March 19 1974 (I had the impresssion it was in January or February but they both seem to agree on \(\sim\)March 19).

The work with BPHZ where most combinatorial theorems had been proved went so fast that I did not want to sign the paper, but they insisted. This was PL \textbf{52B}, 344 (1974) where Slavnov symmetry is descibed in words and is only applied to prove unitarity in the Abalian case (the hardest piece), thought to be simpler than the non Abelian one. Well, in this stupid renormalization framework, semisimple is easier than Abelian and the assumed charge conjugation invariance saved us from total failure. The inconvenience of \(s\ol\w\ =\) gauge fctn was fixed in Erice 75 by the introduction of the Stueckelberg (Nakanishi Lautrup) multiplier \(s\ol\w=b\ sb=0\) which yields a consistent definition of physics.

Conclusion: stupid renormalization did work in this complicated case, too (to my great surprise!): even if one is not clever enough to invent a good regularization, one has a chance to get somewhere. Which renormalization scheme one uses is irrelevant provided it is a renormalization (subject to local ambiguities consistent with power counting).

This little story has its roots in the Marseille 71-72 meethings and your graphical identities \(+\) Landau \(\ra\) Feynman.

About these meetings I would like to ask you what are your memories about the asymptotic freedom affair. I only remember a discussion at the blackboard between you and Symanzik after the last morning lecture where you were comparing signs between Yang Mills anf \(\vv^4\) (?) -- I do not have with me proceeedings of the 1972 meeting if they exist\fn{Neither has the author; we never saw it.} (they may be somewhere in Marseille). In fact I thought this was in 1971, but 1) I have the proceedings showing you were not there, Tini talked about your work (very short abstract)\fn{The proceedings only gives the title; I (the author) was not at that conference.} \\ 2) Slavnov's preprint did not exist yet, I believe and I am sure I gave you a xerox. So, question: do you have anything written on this sign, or 3) did youmention it in your formal talk, or only at this blackboard discussion with Symanzik.\fn{The results on the sign were only published later\,\cite{GtHrengr}; the remarks were made 
in the discussion session after Symanzik's report in the Marseille conference.}
What other witnesses could we ask -- before it is too late -- ?\\
So, any memory is welcome!\fn{Symanzik mentions in a letter to A.~Zee, with a copy to the author, in particular Th.~Appelquist and J.~Primack, but also C.\,P.~Korthals Altes (who organised the conference), W.~Bardeen, J.~Iliopoulos and J.~Prentki were there.}
\\
My best regards,\\
signed R. Stora.

\end{document}